\documentclass[aps,pra,floatfix,superscriptaddress,twocolumn,showpacs,10pt]{revtex4-1}
\usepackage{amssymb, amsmath,color,mciteplus,graphicx,subfigure}
\usepackage{calc,epsfig,epstopdf,color,mciteplus,bm,mathrsfs}
\usepackage{times}
\usepackage{multirow}
\usepackage{bbm}

\begin{document}

\title{Single-Loop and Composite-Loop Realization of Nonadiabatic Holonomic Quantum Gates\\
 in a Decoherence-free Subspace}

\author{Zhennan Zhu} \email{These two authors contribute equally to this work.}
\affiliation{
Hefei National Laboratory for Physical Sciences at the Microscale and Department of Modern Physics, University of Science and Technology of China, Hefei 230026, China}
\affiliation{
CAS Key Laboratory of Microscale Magnetic Resonance, University of Science and Technology of China, Hefei, Anhui 230026, China}

\author{Tao Chen} \email{These two authors contribute equally to this work.}
\affiliation{Guangdong Provincial Key Laboratory of Quantum Engineering and Quantum Materials, GPETR Center for Quantum Precision Measurement, and School of Physics and Telecommunication Engineering, South China Normal University, Guangzhou 510006, China}

\author{Xiaodong Yang}
\affiliation{
Hefei National Laboratory for Physical Sciences at the Microscale and Department of Modern Physics, University of Science and Technology of China, Hefei 230026, China}
\affiliation{
CAS Key Laboratory of Microscale Magnetic Resonance, University of Science and Technology of China, Hefei, Anhui 230026, China}

\author{Ji Bian}
\affiliation{
Hefei National Laboratory for Physical Sciences at the Microscale and Department of Modern Physics, University of Science and Technology of China, Hefei 230026, China}
\affiliation{
CAS Key Laboratory of Microscale Magnetic Resonance, University of Science and Technology of China, Hefei, Anhui 230026, China}

\author{Zheng-Yuan Xue}\email{zyxue83@163.com}
\affiliation{Guangdong Provincial Key Laboratory of Quantum Engineering and Quantum Materials,
GPETR Center for Quantum Precision Measurement,
and School of Physics and Telecommunication Engineering,
South China Normal University, Guangzhou 510006, China}

\author{Xinhua Peng}\email{xhpeng@ustc.edu.cn}
\affiliation{
Hefei National Laboratory for Physical Sciences at the Microscale and Department of Modern Physics, University of Science and Technology of China, Hefei 230026, China}
\affiliation{
CAS Key Laboratory of Microscale Magnetic Resonance, University of Science and Technology of China, Hefei, Anhui 230026, China}
\affiliation{
Synergetic Innovation Center of Quantum Information and Quantum Physics, University of Science and Technology of China, Hefei, Anhui 230026, China}
\affiliation{
College of Physics and Electronic Science, Hubei Normal University, Huangshi, Hubei 435002, China}

\date{\today}

\begin{abstract}
High-fidelity quantum gates are essential for large scale quantum computation, which can naturally be realized in a noise resilient way. It is well-known that geometric manipulation and decoherence-free subspace encoding are promising ways towards robust quantum computation. Here, by combining the advantages of both strategies, we propose and experimentally realize universal holonomic quantum gates in both a single-loop and composite scheme, based on nonadiabatic and non-Abelian geometric phases, in a decoherence-free subspace with nuclear magnetic resonance. Our experiment only employs two-body resonant spin-spin interactions and thus is experimental friendly. In particularly, we also experimentally verify that the composite scheme is more robust against the pulse errors over the single-loop scheme. Therefore, our experiment provides a promising way towards faithful and robust geometric quantum manipulation.
\end{abstract}

\maketitle

\section{Introduction}
It is generally believed that quantum computers can be more efficient in processing certain hard tasks, which cannot be achievable by their classical counterparts. However, quantum information is very fragile and can be destroyed by the weak environmental induced noises. Meanwhile, imperfect quantum manipulation will also introduce additional errors. Therefore, to obtain high-fidelity quantum manipulation, it is essential to fight against various noises and operation errors.

As it is well known, geometric phases~\cite{berry,b3,b2} have some built-in noise-resilient feature \cite{ps1,zhu05,jtt,ps2,mj}, which are determined by the global properties of the evolution paths. Therefore, geometric quantum computation \cite{gqc}, where quantum gates are induced by geometric transformations, is a promising candidate to achieve high-fidelity quantum manipulation. Moreover,  due to the intrinsic noncommutativity, non-Abelian geometric phases \cite{b3} can naturally lead to universal quantum gates, i.e., the so-called holonomic quantum computation \cite{zanardi,adiabatic,duan,adiabatic6}. However,  geometric phases based on adiabatic evolutions are
so slow that decoherence will introduce considerable gate errors \cite{xbwang,zhu}. To deal with this difficulty, nonadiabatic holonomic quantum computation (NHQC) has been proposed recently \cite{Sjoqvist2012, Xu2015, Herterich2016,Xue2017}, where fast holonomic quantum gates can be obtained based on nonadiabatic non-Abelian geometric phases.
In addition, elementary quantum operations of NHQC have also been experimentally demonstrated in nuclear magnetic resonance (NMR) \cite{Feng2013,li2017}, superconducting circuits \cite{Abdumalikov2013,xuy2018,ibm}, and  electron spins in diamond \cite{Zu2014, Arroyo-Camejo2014, nv2017, nv20172,nv20181,nv20182}.
An alternative approach against decoherence is to utilize decoherence-free subspace (DFS) encoding \cite{DFS1,DFS2,DFS3}. Recently, many efforts have also been made to combine NHQC with  DFS encoding \cite{xu2012,n3,n4,xue2014,xue2015a,xue2015b,xue2016,zhao2017}, which can maintain both the noise resilience of the encoding and the operational robustness of holonomies. However, these schemes generally involve three-body or dispersively induced interactions, which are rather complicated and thus difficult to implement experimentally.

Here, we propose and experimentally realize an  NHQC scheme in a three-qubit DFS \cite{xue2015b,xue2016}, based on the resonant single-loop scenario \cite{xue2018}.  Therefore, comparing with previous schemes \cite{xue2015b,xue2016}, our implementation simplifies the needed gate sequences for large-scale algorithm, as it can achieve an arbitrary gate in a single step. The other distinct merit of our proposal is that it only involves resonant two-body interactions of two-level systems, thus leading to fast NHQC in a simplified setup.  However, the robustness against systematic errors of the single-loop implementation is still the same as previous schemes. Then, we move another step further to  incorporate  the composite-loop technique \cite{composite,xue2018b} into our implementation,  which is achieved by changing the way of accumulating the geometric phase. In addition, both the the single-loop and composite-loop implementations are experimentally tested, our experimental comparison between the two  implementations shows that the composite-loop one can indeed further improve the noise resilience of the implemented holonomic quantum gates. Finally, we want to emphasize that all the DFS encoding, the single-loop and the composite-loop strategies have not yet been experimentally demonstrated. Therefore, our experiment  provides a promising methodology towards robust geometric quantum computation.

\section{Single-loop and composite NHQC in a DFS}
To realize NHQC in DFS, three physical qubits are encoded as a logical qubit. This DFS is thus spanned by the single-excitation vectors: $S_1=\{|100\rangle, |001\rangle, |010\rangle \}=\{|0\rangle_L, |1\rangle_L, |E\rangle_L\}$, where a natural encoding of the logical qubit $ | \psi\rangle_L = a |0\rangle_L  + b|1\rangle_L$ and $|E\rangle_L$ is an ancillary state of the logical qubit; $|mnk\rangle\equiv|m\rangle_1\otimes|n\rangle_2\otimes|k\rangle_3$ with the subscript indicating different physical qubits $(q_1, q_2, q_3)$.

\begin{figure}[tb]
\includegraphics[width=\columnwidth]{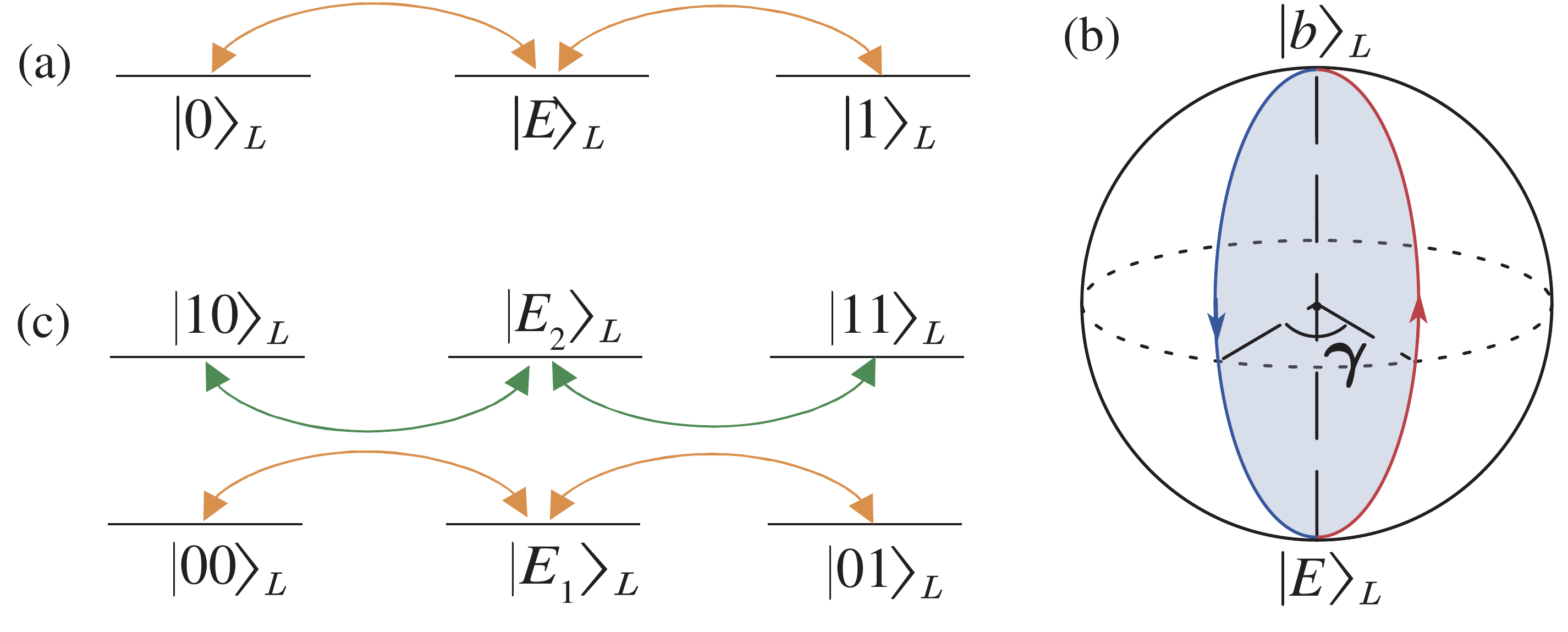}
\caption{Illustration of the proposed setup of our scheme. Effective coupling diagrams for three physical qubits used to realize (a) universal single-logical qubit gates and (c) nontrivial two-logical qubit holonomic gates. (b) Geometrical illustration of single-logical qubit gates by the orange-slice-shaped path.}
  \label{Figure1}
\end{figure}

\subsection{Universal single-qubit gates}
Firstly, we proceed to introduce the construction of universal single-logical-qubit holonomic gates. In order to realize the dynamic construction of the effective $\Lambda$-type Hamiltonian based on DFS encoding \cite{xue2015b,xue2016}, according to the resonant coupling form between physical qubits, the interaction Hamiltonian we design is $\mathcal{H}_S=\mathcal{H}_{1}(\Omega_{1},\phi_1) +\mathcal{H}_{2}(\Omega_{2},\phi_2) $ with
\begin{eqnarray}
\label{EqHS}
\mathcal{H}_{i} (\Omega_{i},\phi_i) &=&\frac {\Omega_{i}} {2}[ \cos\phi_i(X_iX_{i+1}+Y_iY_{i+1})\notag \\
&&+ (-1)^{i+1} \sin\phi_i(X_iY_{i+1}-Y_iX_{i+1})],
\end{eqnarray}
where $i=1, 2$ and $\mathcal{H}_{i} (\Omega_{i},\phi_i) $ denotes the interaction Hamiltonian between the $q_i$ and $q_{i+1}$ physical qubits with the the strength $\Omega_i$ and the phase  $\phi_i$; $X_i, Y_i$, and $Z_i$ denote the Pauli operators for the physical qubit $q_i$.

Setting $\Omega_{1}=\Omega \cos (\theta/2)$, $\Omega_{2}=\Omega \sin(\theta/2)$ with $\Omega=\sqrt{\Omega_{1}^2+\Omega_{2}^2}$ and $\theta=2\tan^{-1}(\Omega_{2}/\Omega_{1})$, as shown in Fig. \ref{Figure1}(a), the Hamiltonian $\mathcal{H}_S$ in the DFS $S_1$ can be written as
\begin{eqnarray}
\label{EqHLS}
\mathcal{H}_{S}&=&\Omega e^{i \phi_1} \left(\cos\frac {\theta} {2} |0\rangle_L+\sin\frac {\theta} {2} e^{i \phi}|1\rangle_L\right)\langle E|+ \mathrm{H.c.}, \notag \\
&=&\Omega e^{i \phi_1} |b\rangle_L \langle E|+ \mathrm{H.c.},
\end{eqnarray}
where $|b\rangle_L=\cos(\theta/2)|0\rangle_L+\sin(\theta/2)e^{i \phi}|1\rangle_L$ with $\phi=\phi_2-\phi_1$.
In the dressed-state representation $\{|b\rangle_L, |d\rangle_L, |E\rangle_L \}$, the dynamic process of the Hamiltonian $\mathcal{H}_{S}$ can be regarded as a resonant coupling between the bright state $|b\rangle_L$ and the ancillary state $|E\rangle_L$, while the dark state $|d\rangle_L=\sin(\theta/2)|0\rangle_L-\cos(\theta/2)e^{i \phi}|1\rangle_L$ decouples from the dynamics all the time.

Thereafter, an arbitrary single-logical-qubit holonomic gate in $S_1$ can be realized with a single-loop scenario, by engineering the quantum system to evolve along an orange-slice-shaped path, as shown in Fig. \ref{Figure1}(b). In our construction, the evolution area is set as $\Omega \tau =\pi$, with $\tau$ being the entire evolution time, which is separated into two equal segments. In the second segment $[0, \tau/2]$, we set $\phi_1=0$, then $\mathcal{H}_{S}$ is reduced to $\mathcal{H}_a=\Omega(|b\rangle_L \langle E|+|E\rangle_L \langle b|)$ and the corresponding evolution operator is $U_a=|d\rangle_L \langle d|-i(|b\rangle_L \langle E|+|E\rangle_L \langle b|)$. In the first segment $[\tau/2, \tau]$, we change the phase $\phi_1$ to $\phi_1'=\pi+\gamma$, then $\mathcal{H}_{S} = \mathcal{H}_b =-\Omega(e^{i \gamma}|b\rangle_L \langle E|+e^{-i \gamma}|E\rangle_L \langle b|)$ and the corresponding evolution operator $U_b^{\gamma} =|d\rangle_L \langle d|+i(e^{i \gamma}|b\rangle_L \langle E|+e^{-i \gamma}|E\rangle_L \langle b|)$. In this way, in the logical-qubit computational basis $\{|0\rangle_L, |1\rangle_L\}$, the induced gate operation will be    
\begin{eqnarray}\label{EqUS}
&&U_S(\gamma, \theta, \phi) =U_b^{\gamma} U_a
=|d\rangle_L\langle d|+e^{i\gamma}|b\rangle_L\langle b| \notag\\
&=&e^{i\frac {\gamma} {2}} \left(
               \begin{array}{cccc}
                \cos\frac {\gamma} {2}+i\sin\frac {\gamma}{2} \cos\theta & i\sin\frac {\gamma}{2} \sin\theta e^{-i\phi} \\
                 i\sin\frac {\gamma}{2} \sin\theta e^{i\phi} &  \cos\frac {\gamma} {2}-i\sin\frac {\gamma}{2} \cos\theta
               \end{array}
               \right)\notag\\
 &=&e^{i\frac {\gamma} {2}} e^{i\frac {\gamma} {2} \vec{n}\cdot\vec{\sigma}_L},
\end{eqnarray}
where $\vec{\sigma}_{_L}=(X^L, Y^L, Z^L)$ are the Pauli operators for the logical qubit and  $\vec{n}=(\sin{\theta}\cos{\phi},\sin{\theta}\sin{\phi},\cos{\theta})$. In the Bloch sphere representation, Eq. (\ref{EqUS})  indicates a rotation operation around the axis $\vec{n}$ by an angle $\gamma/ 2$, up to a global phase factor, which can lead to arbitrary single-logical-qubit gates as both $\vec{n}$ and $\gamma$ are tunable.
In addition, the implemented gates are geometric as the evolution of logical qubit states satisfies ($i$) the parallel-transport condition, i.e., $_L\langle j(t)|\mathcal{H}_{S}|k(t)\rangle_L=0$ with $j,k\in \{b, d\}$, and ($ii$) the cyclic evolution condition, i.e., $|b(\tau)\rangle_L=U_S(\gamma, \theta, \phi)|b\rangle_L=e^{i\gamma}|b\rangle_L$ and
$|d(\tau)\rangle_L=U_S(\gamma, \theta, \phi)|d\rangle_L=|d\rangle_L$.

Usually, the existence of systematic errors tends to devastate the advantage of the robustness of holonomic gate in the NHQC  \cite{zheng,jing}. To overcome this, we suggest  implementing the holonomic gates with composite schemes \cite{composite,xue2018b}. To 
achieve this in DFS, we take $U_S(\gamma/N, \theta, \phi)$ as an elementary gate, where $N>1$. Thus, the target gate $U_S({\gamma} , \theta, \phi)$  in Eq. (\ref{EqUS}) can be achieved by sequentially apply $N$ times of the elementary gate, while keeping the cumulative geometric phase to be $\gamma$, i.e.,
\begin{eqnarray}
\label{UC}
\left[U_S(\gamma/N, \theta, \phi)\right]^N 
=U_S({\gamma}, \theta, \phi).
\end{eqnarray}

\subsection{Nontrivial two-qubit gates}
We now proceed to the construction of nontrivial two-logical-qubit holonomic gates, combining with the above arbitrary single-logical-qubit holonomic gates. For the two-logical qubit, a six-dimensional DFS exists, i.e.,
\begin{eqnarray}
\label{EqS2}
S_2&=&\{|00\rangle_L, |01\rangle_L, |10\rangle_L, |11\rangle_L, |E_1\rangle_L, |E_2\rangle_L\}\notag\\
   &=&\{|100100\rangle, |100001\rangle, |001100\rangle, \notag\\
   & & \ \ |001001\rangle, |101000\rangle, |000101\rangle\},
\end{eqnarray}
where $|E_1\rangle_L$ and $|E_2\rangle_L$ are the ancillary states;  $|mnkm'n'k'\rangle=|m\rangle_1\otimes|n\rangle_2\otimes|k\rangle_3 \otimes|m'\rangle_4 \otimes|n'\rangle_5\otimes|k'\rangle_6$, i.e., the  physical qubits $(q_1, q_2, q_3)$ and  $(q_4, q_5, q_6)$ encode the first and second logical qubits, respectively. For two-qubit case, we design Hamiltonian $\mathcal{H}_T=\mathcal{H}_{3}+\mathcal{H}_{4}$ with
\begin{eqnarray}
\label{EqHT}
\mathcal{H}_{3}&=&\frac {\Omega_{3}} {2}[\cos\varphi(X_3X_4+Y_3Y_4)+ \sin\varphi(Y_3X_4-X_3Y_4)],\notag\\
\mathcal{H}_{4}&=&\frac {\Omega_{4}} {2}  (X_3X_6+Y_3Y_6).
\end{eqnarray}
Defining $\Omega_{3}=\Omega' \cos(\vartheta/2)$, $\Omega_{4}=\Omega' \sin(\vartheta/2)$ with $\Omega'=\sqrt{\Omega_{3}^2+\Omega_{4}^2}$ and $\vartheta=2\tan^{-1}(\Omega_{4}/\Omega_{3})$,   $\mathcal{H}_T$ can be rewritten, in the DFS $S_2$, as $\mathcal{H}_{T}=\mathcal{H}_{LT}^{(1)}+\mathcal{H}_{LT}^{(2)}$, with
\begin{eqnarray}
\label{EqHLT}
\mathcal{H}_{LT}^{(1)}&=&\Omega' \left(e^{-i \varphi}\cos\frac {\vartheta} {2} |00\rangle_L +\sin\frac {\vartheta} {2}|01\rangle_L\right)\langle E_1|+ \mathrm{H.c.}, \notag \\
\mathcal{H}_{LT}^{(2)}&=&\Omega' \left(e^{ i \varphi}\cos\frac {\vartheta} {2} |11\rangle_L +\sin\frac {\vartheta} {2}|10\rangle_L\right)\langle E_2|+ \mathrm{H.c.}, \notag
\end{eqnarray}
being two commuting parts. In the subspace $\{|00\rangle_L, |01\rangle_L, |E_1\rangle_L\}$ or $\{|10\rangle_L, |11\rangle_L, |E_2\rangle_L \}$), $\mathcal{H}_{LT}^{(1)}$ or $\mathcal{H}_{LT}^{(2)}$ forms a Hamiltonian that is similar to $\mathcal{H}_{S}$ in Eq. \eqref{EqHLS} for the single-logical qubit gates, and the two subspaces evolve independently with the coupling diagram, as shown in Fig. \ref{Figure1}(c). When $\Omega'T=\pi$ with $T$ being the evolution time, the evolution operator in $S_2$ is
\begin{eqnarray}
\label{EqUT}
U_T(& &\vartheta, \varphi)=-\left(
\begin{array}{cccc}
\cos\vartheta & \sin\vartheta e^{i \varphi} & 0 & 0 \\
\sin\vartheta e^{-i \varphi} & -\cos\vartheta & 0 & 0 \\
 0 & 0 & -\cos\vartheta & \sin\vartheta e^{i \varphi} \\
 0 & 0 & \sin\vartheta e^{-i \varphi} & \cos\vartheta \\
\end{array}
\right).\notag\\
\end{eqnarray}
As the evolution in the subspace $\{|00\rangle_L, |01\rangle_L \}$ is different from that of in the subspace $\{|10\rangle_L, |11\rangle_L \}$  in general, Eq. (\ref{EqUT}) denotes nontrivial two-qubit gates, by setting deferent $\vartheta$ and/or $\varphi$. For example, a controlled-Z gate ($U_{\textrm{CZ}}$) can be constructed by
\begin{eqnarray}
U_{\textrm{CZ}}=U^2_S\left(\frac{\pi}{2}, \frac{\pi}{2}, \pi\right) \textrm{K}
U^2_S \left(\frac{\pi}{2}, \frac{\pi}{2}, 0\right),
\end{eqnarray}
with
\begin{eqnarray}
\textrm{K}=U^1_S \left(\pi, \frac{\pi}{2}, 0\right) U^2_S \left(\pi, \frac{\pi}{4}, 0\right) U_T \left(\frac{\pi}{4}, 0\right),
\end{eqnarray}
where  superscripts ``1'' and ``2'' label the two logical qubits.

\begin{figure}[tb]
\includegraphics[width=\columnwidth]{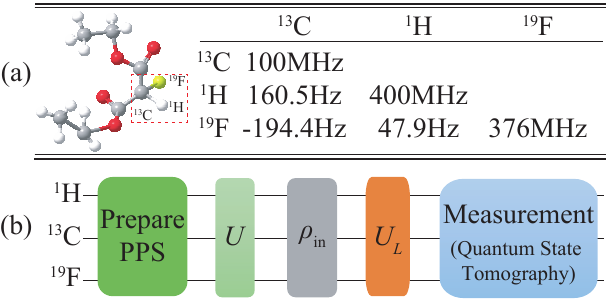}
  \caption{(a) Molecular structure and relevant parameters of Diethyl fluoromalonate. The chemical shifts and scalar couplings are on and below the diagonal of the table, respectively. (b) Experimental scheme for QPT for different holonomic gate $U_L$.}
  \label{Figure2}
\end{figure}

\section{Experimental realizations}
We employ diethyl fluoromalonate dissolved in $^2$H-labeled chloroform at 303K as an NMR quantum simulator, where three physical qubits $(q_1, q_2, q_3)$ are realized by the nuclear spins $(\rm{\,^1H, \,^{13}C, \,^{19}F})$, respectively. The molecular structure and parameters are shown in Fig. \ref{Figure2}(a). The natural Hamiltonian in the triple-resonance rotating frame is
\begin{eqnarray}
\label{GNMR}
\mathcal{H}_{\emph{NMR}}=\frac {\pi}{2} \sum_{1\leq i< j\leq 3}J_{ij}Z_iZ_j,
\end{eqnarray}
where $J_{ij}$ is the scalar coupling strength between the $i$th and $j$th nucleus. The experiment begins with preparing a pseudopure state $\rho_{pps}=(1-\varepsilon) I/8+\varepsilon |000\rangle\langle 000|$
from the thermal equilibrium state, using the line-selective method \cite{Peng2001}. Here, $\varepsilon\approx 10^{-5}$ denotes the polarization, and $I$ denotes the 8$\times$8 identity matrix. Thereafter, the DFS encoded logical states can be obtained by the rotations $R_x^1(\pi)|000\rangle=|100\rangle\equiv|0\rangle_L$,
$R_x^3(\pi)|000\rangle=|001\rangle\equiv|1\rangle_L$. 

In the following, we take holonomic NOT and Hadamard (H) gates as two typical examples of single-logical-qubit gates to experimentally demonstrate their performance. Without loss of generalization, we set $\phi=\phi_2-\phi_1=0$. According to Eq. (\ref{EqUS}), one can obtain NOT $= U_S(\pi, \pi/2, 0 )$ under the evolution of
\begin{eqnarray}
\mathcal{H}_S^N &=& \mathcal{H}_a^N  = \mathcal{H}_b^N  \notag\\
& =& {\sqrt{2} \Omega \over 4}[(X_1X_2+Y_1Y_2+X_2X_3+Y_2Y_3)],
\end{eqnarray}
with duration $\tau = \pi/ \Omega$,  and H $= U_S(\pi, \pi/4, 0 )$  under the evolution of
\begin{eqnarray}
\mathcal{H}_S^H &=& \mathcal{H}_a^H  = \mathcal{H}_b^H  \notag\\
& =& {\Omega \over 2}\cos\left({\pi\over 8}\right)(X_1X_2+Y_1Y_2)\notag\\
&& +{\Omega \over 2} \sin\left({\pi\over 8}\right)(X_2X_3+Y_2Y_3),
\end{eqnarray}
with duration $\tau$, in a single-loop way. Similarly, according to Eq. (\ref{UC}), composite-pulse implementations are NOT $= \left[U_S(\pi/2, \pi/2, 0 )\right]^2 $ with
\begin{eqnarray}
\mathcal{H}_a^{2N} = {\sqrt{2} \Omega \over 4}[(Y_1Y_2+X_1X_2+X_2X_3+Y_2Y_3)],\notag\\
\mathcal{H}_b^{2N} = {\sqrt{2} \Omega \over 4}[(Y_1X_2-X_1Y_2+ X_2Y_3-Y_2X_3)],
\end{eqnarray}
and  $H=  \left[U_S(\pi/2, \pi/4, 0 )\right]^2 $ with
\begin{eqnarray}
\mathcal{H}_a^{2H} &=& {\Omega \over 2}\cos\left({\pi\over 8}\right)(Y_1Y_2 +X_1X_2)\notag\\
&&+{\Omega \over 2}\sin\left({\pi\over 8}\right)(X_2X_3+Y_2Y_3)],\notag\\
\mathcal{H}_b^{2H} &=& {\Omega \over 2}\cos\left({\pi\over 8}\right)(Y_1X_2-X_1Y_2)\notag\\
&&+{\Omega \over 2} \sin\left({\pi\over 8}\right)(X_2Y_3-Y_2X_3),
\end{eqnarray}
for $n = 2$. For the sake of simplicity, we take effective coupling parameter $\Omega=1$ in the Hamiltonian $\mathcal{H}_S$ hereafter. Using Trotter formula, we approximately generate the evolution operator
\begin{eqnarray}\label{Trotter}
e^{-i\mathcal{H}_S \tau}\cong\left(e^{-i \mathcal{H}_{2} {\tau\over 6} }e^{-i\mathcal{H}_{1} {\tau\over 3}}
e^{-i\mathcal{H}_{2} {\tau\over 6}}\right)^3 +O\left[\left({\tau\over 3}\right)^3\right].
\end{eqnarray}
All the gate fidelities can reach 0.9999 by the Trotter approximations, and the corresponding pulse sequences are presented in Appendix A.

\begin{figure}[tb]
\includegraphics[width=\columnwidth]{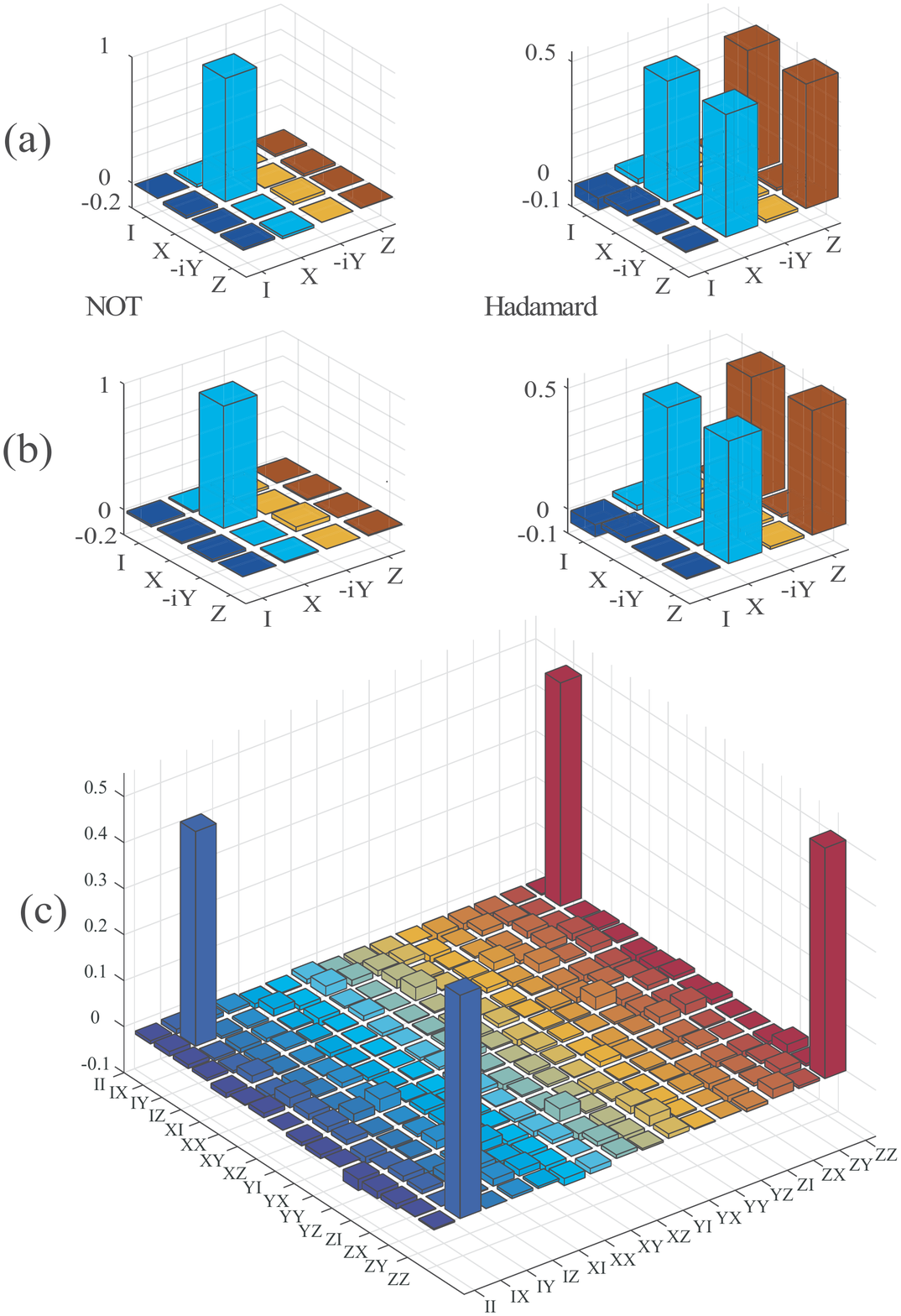}
  \caption{ Experimentally reconstructed $\chi$ matrixes in the logical-qubit subspace for holonomic gates:  (a)  NOT (left) and H (right) gates in the single-loop way, i.e.,  NOT $=U_S(\pi, \pi/2, 0)$  and H $= U_S(\pi,  \pi/4, 0)$. (b)  NOT (left) and H (right) gates in the composite way with $N =2$, i.e.,  NOT $=\left[U_S(\pi/2, \pi/2, 0 )\right]^2$  and H $= \left[U_S(\pi/2, \pi/4, 0)\right]^2$.  (c) A universal two-qubit gate $U_T(\pi/4, 0)$. All the imaginary parts of these $\chi$ matrixes are less than 0.1, and not shown here.} \label{gate}
\end{figure}

In order to quantitatively access experimental implementations of the NHQC gates, we use standard quantum process tomography (QPT) \cite{Chuang1997} in the logical qubit subspace, and the experimental scheme is shown in Fig. \ref{Figure2}(b), see Appendix B for the details. For single-logical-qubit gates, we prepare the initial state $\rho_{in}$ as $|0\rangle_L, |1\rangle_L, (|0\rangle_L+|1\rangle_L)/\sqrt{2}$ and $(|0\rangle_L+i|1\rangle_L)/\sqrt{2}$ through the operation $U$, and then perform holonomic operation $U_L$ for different logical gates, e.g., NOT or H, finally the output state $\rho_{f} = U_L \rho_{in} U_L^{\dagger}$ are determined by quantum state tomography \cite{JS2002}. The required information are selected to reconstruct quantum channels in the logical-qubit subspace.  The experimentally reconstructed $\chi$ matrixes in the logical-qubit subspace for holonomic NOT and H gates are shown in Fig. \ref{gate} for (a) the single-loop way and (b) the composite way. Here, we estimate the quality of the reconstructed gates by the distance of the experimental and theoretic $\chi$ matrixes under the definition of Frobenius-norm \cite{Wang2011}, i.e., $D(\chi)\equiv D(\chi_{exp},\chi_{th})=\|\chi_{exp}-\chi_{th}\|$.
The results are 0.202 and 0.217 for holonomic NOT and H gates in a single-loop way, respectively; 0.216 and 0.210 for those in the composite scheme. These errors mainly come from the imperfection of state preparation and measurement, see Appendix C for details.

For the realization of the two-logical-qubit gates, one finds that only three physical qubits $(q_3, q_4, q_6)$ are active in the Hamiltonian $\mathcal{H}_T$. Therefore the dynamics of the two-logical-qubit gates can be simulated on the three-qubit quantum processor. Neglecting the three uninvolved physical qubits, the reduced two-logical-qubit states are
\begin{eqnarray}
& &|00\rangle_L\Rightarrow |010\rangle_{346}, \quad|01\rangle_L\Rightarrow |001\rangle_{346},\notag \\
& &|10\rangle_L\Rightarrow |110\rangle_{346}, \quad|11\rangle_L\Rightarrow |101\rangle_{346},\notag \\
& &|E_1\rangle_L\Rightarrow|100\rangle_{346}, \quad|E_2\rangle_L\Rightarrow|011\rangle_{346}.
\label{EqSUBSPACE}
\end{eqnarray}
In our experiment, the nuclear spins $(\rm{\,^1H, \,^{13}C, \,^{19}F})$ are chosen as physical qubits $(q_4, q_3, q_6)$. Similar to the case of single-logical-qubit gate, a two-logical-qubit gate $U_T(\pi/4, 0)$ can also be implemented under the evolution of $\mathcal{H}_T =\Omega'[\cos(\pi/8)(X_3X_4+Y_3Y_4)+\sin(\pi/8)(X_3X_6+Y_3Y_6)]/2$ with duration $T = \pi/ \Omega'$, where $\Omega'=1$ for simplicity. Likely, we perform the standard QPT for two-logical-qubit gates in the logical-qubit subspace, by preparing 16 initial states
$ \{|0\rangle_L,|1\rangle_L,(|0\rangle_L+|1\rangle_L)/\sqrt{2}, (|0\rangle_L+i|1\rangle_L)/\sqrt{2}\} \otimes\{|0\rangle_L,|1\rangle_L,(|0\rangle_L+|1\rangle_L)/\sqrt{2}, (|0\rangle_L+i|1\rangle_L)/\sqrt{2}\}.$
Therefore, the $\chi$ matrix for the two-logical-qubit gate $U_T(\pi/4, 0)$ can be experimentally determined, as shown in Fig. \ref{gate}(c), and the gate distance between the experimental and theoretical ones is $0.274$.

\begin{figure}[tb]
\includegraphics[width=8cm]{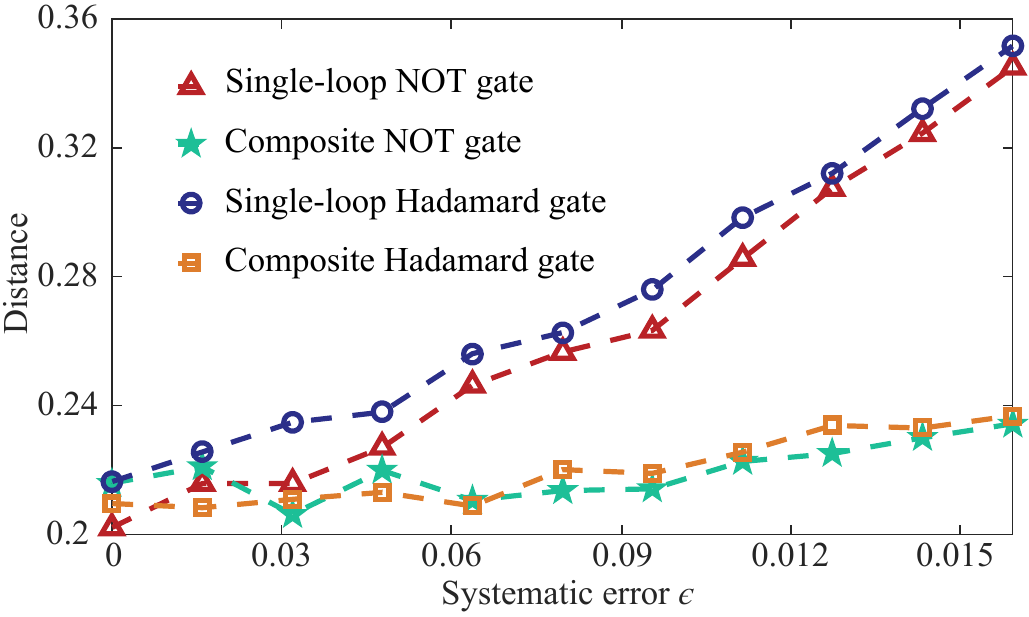}
  \caption{ Experimental gate distances with respective to systematic error $\epsilon$ for single-logical-qubit holonomic NOT and H gates in the single-loop way and in the composite way with $N = 2$.}
\label{Figure4}
\end{figure}

\section{Robustness test}
In the following, we shall experimentally test the robustness of nonadiabatic holonomic quantum gates by taking the single-logical-qubit gates as examples. To do this, we add systematic errors in Hamiltonian $\mathcal{H}_S$ as $(1+\epsilon) \Omega$ with $\epsilon$ being the error fraction, i.e., the deviation of coupling strength. This might be caused by the imperfection of $\pi$-evolution condition so that the cyclic evolution is no longer satisfied.
Using the same QPT procedure above, we obtain the gate distances versus the error fraction $\epsilon$ for nonadiabatic holonomic quantum gates in both the single-loop and composite schemes, shown in Fig. \ref{Figure4}. This result indicates that holonomic gates realized by the composite scheme have better robustness against the systematic error $\epsilon$. The abnormal behaviors in the small-systematic-error range for NOT gate are mainly due to the imperfection of the state preparation and measurement, which dominates the main errors  when $\epsilon$ is small. In addition, we note that the gate infidelity induced by the initial state preparation can be further suppressed \cite{initial}.

\section{Summary}
By combining the advantages of geometric manipulation and DFS encoding, we have proposed an extended NQHC scheme, and demonstrated its feasibility in a proof-of-principle experiments via an NMR quantum information processor. We experimental demonstrate universal NHQC in DFS for both the single-loop and composite way, which is an important step-toward for fault-tolerant quantum computing. Moreover, we also test the robustness of our implemented gates and show that the holonomic gates realized in the composite way does have a better performance against the systematic error than in the single-loop case.

\acknowledgements

This work was supported by National Key Research and Development Program of China (Grants No. 2018YFA0306600 and No. 2016YFA0301803),
National Natural Science Foundation of China (Grants No. 11425523,  No. 11661161018, and No. 11874156), the
Key R\&D Program of Guangdong Province (Grant No. 2018B030326001),
and Anhui Initiative in Quantum Information Technologies (Grant No. AHY050000).

\begin{figure}[tb]
\includegraphics[width=8cm]{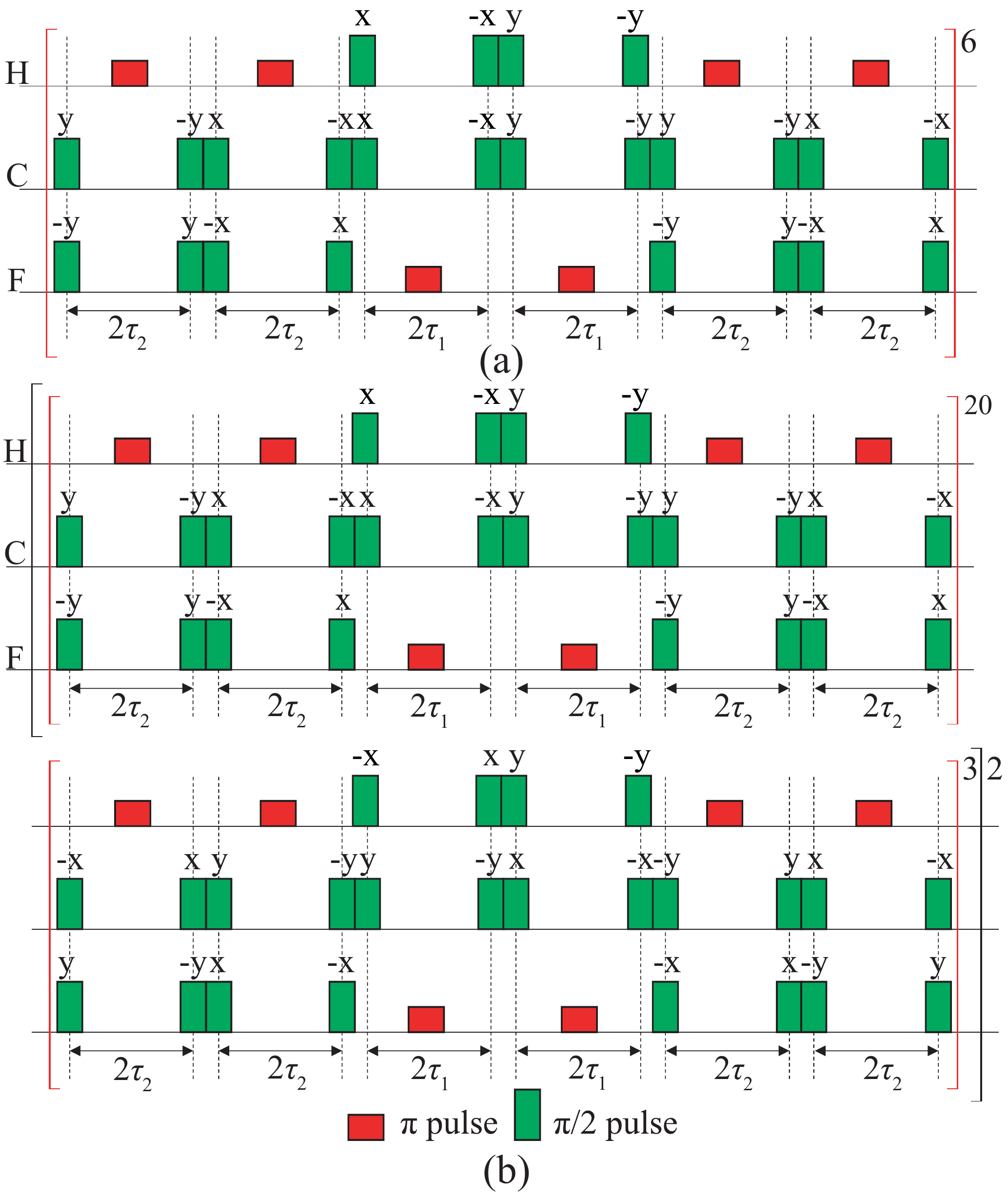}
  \caption{Experimental pulse sequences for the single-logical-qubit gates $U_S(\pi, \theta, 0)$ realized (a)  in a single loop and (b)  in a composite scheme with $N =2$. The free evolution time $\tau_1= \cos\frac{\theta}{2}/(12J_{\rm{CH}})$, $\tau_2= \sin\frac{\theta}{2}/(24 J_{\rm{CF}})$ .}
\label{FigureS2}
\end{figure}

\begin{figure*}[tb]
\includegraphics[width=\linewidth]{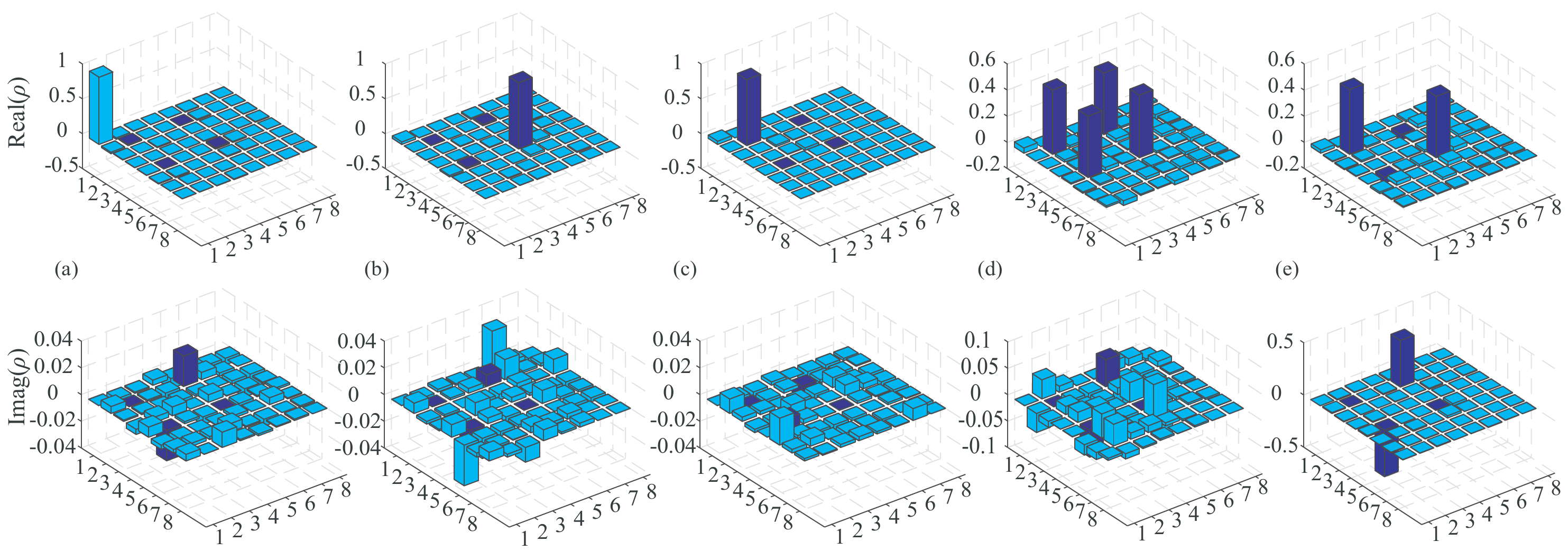}
  \caption{Experimentally reconstructed density matrices of initial states for single-logical-qubit gates:  (a) - (e) respectively correspond to $|000\rangle, |0\rangle_L, |1\rangle_L, (|0\rangle_L+|1\rangle_L)/\sqrt{2},(|0\rangle_L+i|1\rangle_L)/\sqrt{2}$, where the elements in the logical-qubit subspace $S_1$ are marked as dark blue bars. }
\label{FigureS3}
\end{figure*}

\begin{figure*}[tb]
\includegraphics[width=0.83\linewidth]{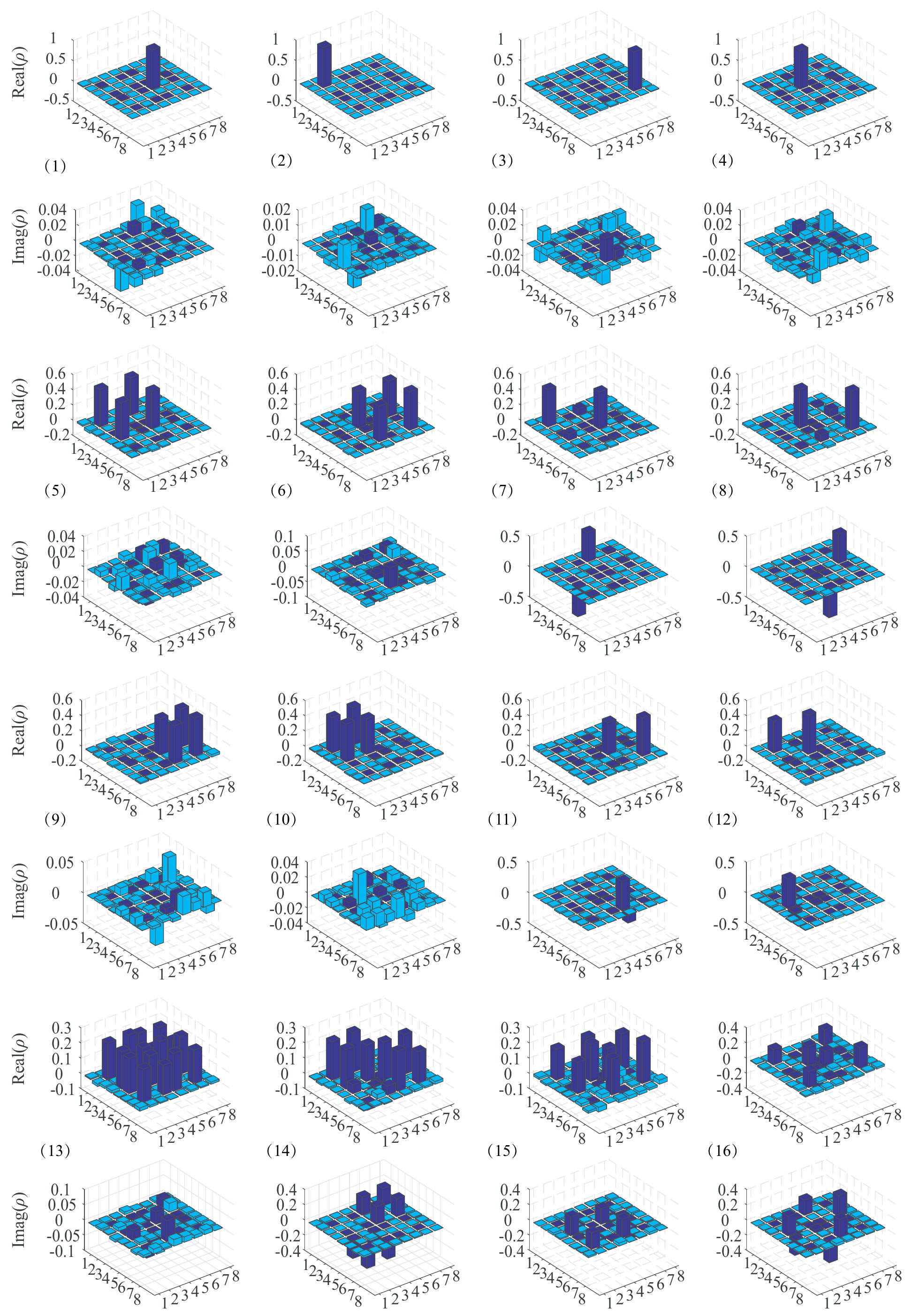}
  \caption{ Experimentally reconstructed density matrices of initial states for two-logical-qubit gates:  (1) to (16) respectively correspond to $|00\rangle_L, |01\rangle_L, |10\rangle_L, |11\rangle_L,
|00\rangle_L+|01\rangle_L)/\sqrt{2}, (|10\rangle_L+|11\rangle_L)/\sqrt{2}, (|00\rangle_L+i|01\rangle_L)/\sqrt{2},  (|10\rangle_L+i|11\rangle_L)/\sqrt{2},
(|00\rangle_L+|10\rangle_L)/\sqrt{2}, (|01\rangle_L+|11\rangle_L)/\sqrt{2}, (|00\rangle_L+i|10\rangle_L)/\sqrt{2}, (|01\rangle_L+i|11\rangle_L)/\sqrt{2},
(|00\rangle_L+|10\rangle_L+|01\rangle_L+|11\rangle_L)/2, (|00\rangle_L+i|01\rangle_L+|10\rangle_L+i|11\rangle_L)/2,
(|00\rangle_L+i|10\rangle_L+|01\rangle_L+i|11\rangle_L)/2, (|00\rangle_L+i|01\rangle_L+i|10\rangle_L-|11\rangle_L)/2$, where the elements in the two-logical-qubit subspace $S_2$ are marked as dark blue bars.}
\label{FigureS4}
\end{figure*}

\appendix

\section{Experimental pulse sequences}

Starting from the Hamiltonian of constructing holonomic quantum gates in Eq. (\ref{EqHS}), expectedly, the target interaction Hamiltonian in experiment we want to design is
\begin{eqnarray}
\mathcal{H}_S(\Omega, \theta, \phi; \phi_1)  &  = &  \mathcal{H}_S(\Omega_1, \phi_1; \Omega_2, \phi_2) = \sum_{i = 1}^{2} \mathcal{H}_i(\Omega_i, \phi_i)  \notag \\
 &  = &  \frac {\Omega} {2} \{ \cos  \frac {\theta}{2} [ \cos\phi_1(X_1X_{2}+Y_1Y_{2 })  \notag \\
 &   & + \sin\phi_1(X_1Y_{2} -Y_1X_{2})  ] \notag\\
  &   & + \sin \frac {\theta}{2} [ \cos(\phi_1 + \phi)(X_2X_{3}+Y_2Y_{3 })  \notag \\
 &   & - \sin(\phi_1 + \phi)(X_2Y_{3} -Y_2X_{3})] \} .
\end{eqnarray}
Then, an arbitrary single-logical-qubit gate
\begin{eqnarray}
U_S(\gamma, \theta, \phi) = e^{-i \mathcal{H}_S(\Omega, \theta, \phi; \gamma + \pi)\frac{ \tau} {2}} e^{-i \mathcal{H}_S(\Omega, \theta, \phi; 0) \frac{ \tau} {2}},
\end{eqnarray}
can be achieved  in a single-loop way, by setting $\Omega=\sqrt{\Omega_{1}^2+\Omega_{2}^2}$, $\theta=2\tan^{-1}(\Omega_{2}/\Omega_{1})$ and $\phi = \phi_2 - \phi_1$, with $\Omega \tau = \pi$. For the NOT and Hadamard gates, the operator has the following form
\begin{eqnarray}
U_S(\pi, \theta, 0) & =&  e^{-i \mathcal{H}_S(\Omega, \theta, 0; 2 \pi)\frac{ \tau} {2}} e^{-i \mathcal{H}_S(\Omega, \theta, 0; 0) \frac{ \tau} {2}} \notag \\
 & =&   e^{-i \mathcal{H}_S(\Omega, \theta, 0;  0) \tau},
\end{eqnarray}
due to the fact that $\mathcal{H}_S(\Omega, \theta, 0; 2\pi) = \mathcal{H}_S(\Omega,\theta,0; 0) = \Omega[\cos(\theta/2)(X_1X_2+Y_1Y_2)+\sin(\theta / 2)(X_2X_3+Y_2Y_3)]/2$.
 The holonomic NOT and H gates correspond to $\theta=\pi/2$ and $\pi/4$, respectively.

Using Trotter formulas in Eq. (\ref{Trotter}), we can design the experimental pulse sequence for the realization of the $U_S(\pi, \theta, 0)$ gate, as shown in Fig. \ref{FigureS2} (a). Similarly, for the realization of a composite gate with $n =2$, $U_S(\pi, \theta, 0) = \left[U_S(\pi/2, \theta, 0 )\right]^2$, where
\begin{eqnarray}
U_S\left(\pi/2, \theta, 0\right) & =&  e^{-i \mathcal{H}_S(\Omega, \theta, 0; \frac{3\pi}{2} ) \frac{ \tau} {2}} e^{-i \mathcal{H}_S(\Omega, \theta, 0; 0) \frac{ \tau} {2}},
\end{eqnarray}
with $ \mathcal{H}_S(\Omega, \theta, 0; 3\pi/2) =  \Omega [ -\cos  \frac {\theta}{2} (X_1Y_{2}  - Y_1X_{2} ) + \sin \frac {\theta}{2} (X_2Y_{3}  - Y_2X_{3} ) ]/2 $. Fig. \ref{FigureS2} (b) shows the whole experimental pulse sequence for $U_S(\pi, \theta, 0) $ in the realization of the  composite gate scheme with $N =2$.

For the experimental realization of the  two-logical-qubit gate  $U_T(\pi/4, 0)$, the target Hamiltonian is $\mathcal{H}_T =\Omega'[\cos(\pi/8)(X_3X_4+Y_3Y_4)+\sin(\pi/8)(X_3X_6+Y_3Y_6)]/2$, which is the same as the Hamiltonian for the holonomic H gate, except for the qubit labeling. Therefore, it can also be implemented by the pulse sequence shown in Fig. \ref{FigureS2}.

\begin{table*}
  \centering
\begin{tabular}{ccccc}
\hline
\hline
Order & Logical qubit & Physical qubit($\rm{^1H, ^{13}C, ^{19}F}$) & $D^I_F$ & $D^I_L$\\
\hline
 {1}& {{$|0\rangle_L$}}& {{$|100\rangle$}} & {{0.070}}& {{0.055}}\\
 {{2}}& {{$|1\rangle_L$}}& {{$|001\rangle$}} & {{0.072}}& {{0.060}} \\
 {{3}}& {{$|+\rangle_L$}}& {{$(|100\rangle+|001\rangle)/\sqrt{2}$}} & {{0.147}}& {{0.114}} \\
 {{4}}& {{$|-\rangle_L$}}& {{$(|100\rangle+i|001\rangle)/\sqrt{2}$}}& {{0.164}}& {{0.149}} \\
 {{5}}& {{$|00\rangle_L$}}& {{$|100\rangle$}} & {{0.070}}& {{0.056}}\\
 {{6}}& {{$|01\rangle_L$}}& {{$|001\rangle$}} & {{0.072}}& {{0.063}}\\
 7&$|10\rangle_L$& $|110\rangle$ & 0.101& 0.077\\
 8&$|11\rangle_L$& $|011\rangle$ & 0.113& 0.086\\
 {{9}}& {{$(|00\rangle_L+|01\rangle_L)/\sqrt{2}$}}& {{$(|100\rangle+|001\rangle)/\sqrt{2}$}} & {{0.147}}& {{0.128}}\\
 10&$(|10\rangle_L+|11\rangle_L)/\sqrt{2}$& $(|110\rangle+|011\rangle)/\sqrt{2}$  & 0.187& 0.170\\
 {{11}}&{{$(|00\rangle_L+i|01\rangle_L)/\sqrt{2}$}}& {{$(|100\rangle+i|001\rangle)/\sqrt{2}$}}& {{0.164}}& {{0.155}}\\
 12&$(|10\rangle_L+i|11\rangle_L)/\sqrt{2}$& $(|110\rangle+i|011\rangle)/\sqrt{2}$& 0.186& 0.164\\
 13&$(|00\rangle_L+|10\rangle_L)/\sqrt{2}$& $(|100\rangle+|110\rangle)/\sqrt{2}$  & 0.118& 0.082\\
 14&$(|01\rangle_L+|11\rangle_L)/\sqrt{2}$& $(|001\rangle+|011\rangle)/\sqrt{2}$  & 0.121& 0.081\\
 15&$(|00\rangle_L+i|10\rangle_L)/\sqrt{2}$& $(|100\rangle+i|110\rangle)/\sqrt{2}$& 0.115& 0.088\\
 16&$(|01\rangle_L+i|11\rangle_L)/\sqrt{2}$& $(|001\rangle+i|011\rangle)/\sqrt{2}$& 0.130& 0.101\\
 17&$(|00\rangle_L+|01\rangle_L+|10\rangle_L+|11\rangle_L)/2$& $(|100\rangle+|001\rangle+|110\rangle+|011\rangle)/2$    & 0.190& 0.169\\
 18&$(|00\rangle_L+i|01\rangle_L+|10\rangle_L+i|01\rangle_L)/2$& $(|100\rangle+i|001\rangle+|110\rangle+i|011\rangle)/2$& 0.152& 0.135\\
 19&$(|00\rangle_L+|01\rangle_L+i|10\rangle_L+i|11\rangle_L)/2$& $(|100\rangle+|001\rangle+i|110\rangle+i|011\rangle)/2$& 0.198& 0.151\\
 20&$(|00\rangle_L+i|01\rangle_L+i|10\rangle_L-|11\rangle_L)/2$& $(|100\rangle+i|001\rangle+i|110\rangle-|011\rangle)/2$& 0.169& 0.144\\
\hline
\hline
\end{tabular}
\caption{Distances of all initial states experimentally reconstructed. Here $D(C) = \|C\| = \sqrt{Tr(C C^{T})}$ is the matrix F-norm of the matrix $C$ defined by $ C =C_{exp}-C_{th}$ to quantify the closeness of the experimental matrix $C_{exp}$ and the idea one $C_{th}$. Here $C^{T}$ is the conjugate transpose of $C$. Subscript $F$ and $L$ denote the three-physical-qubit space and the logical-qubit subspace.}\label{SF}
\end{table*}

\begin{table*}
  \centering
\begin{tabular}{cccccccccccc}
\hline
\hline
Order & Initial states & $D^{X1}_F$ & $D^{X1}_L$& $D^{X2}_F$ & $D^{X2}_L$& $ D^{H1}_F$ & $D^{H1}_L$& $ D^{H2}_F$ & $D^{H2}_L$ & $D^2_F$ & $D^2_L$\\
\hline
 1&$ |0\rangle_L$& 0.243& 0.187& 0.244& 0.171& 0.289& 0.198& 0.259& 0.184& & \\
 2&$ |1\rangle_L$& 0.275& 0.191& 0.228& 0.165& 0.285& 0.191& 0.251& 0.185& & \\
 3&$ |+\rangle_L$& 0.305& 0.225& 0.287& 0.237& 0.319& 0.233& 0.278& 0.218& & \\
 4&$ |-\rangle_L$& 0.259& 0.160& 0.278& 0.203& 0.254& 0.195& 0.250& 0.179& & \\
 5&$|00\rangle_L$&      &      & & & & & & & 0.289& 0.221 \\
 6&$|01\rangle_L$&      &      & & & & & & & 0.279& 0.231 \\
 7&$|10\rangle_L$&      &      & & & & & & & 0.278& 0.249 \\
 8&$|11\rangle_L$&      &      & & & & & & & 0.256& 0.228 \\
 9&$(|00\rangle_L+|01\rangle_L)/\sqrt{2}$& & & & & & & & &  0.307& 0.255 \\
 10&$(|10\rangle_L+|11\rangle_L)/\sqrt{2}$& & & & & & & & & 0.268& 0.235 \\
 11&$(|00\rangle_L+i|01\rangle_L)/\sqrt{2}$& & & & & & & & &0.236& 0.208 \\
 12&$(|10\rangle_L+i|11\rangle_L)/\sqrt{2}$& & & & & & & & &0.242& 0.215 \\
 13&$(|00\rangle_L+|10\rangle_L)/\sqrt{2}$& & & & & & & & & 0.286& 0.243 \\
 14&$(|01\rangle_L+|11\rangle_L)/\sqrt{2}$& & & & & & & & & 0.263& 0.210 \\
 15&$(|00\rangle_L+i|10\rangle_L)/\sqrt{2}$& & & & & & & & &0.250& 0.214 \\
 16&$(|01\rangle_L+i|11\rangle_L)/\sqrt{2}$& & & & & & & & &0.231& 0.171\\
 17&$(|00\rangle_L+|01\rangle_L+|10\rangle_L+|11\rangle_L)/2$& & & & & & & &   & 0.276& 0.212 \\
 18&$(|00\rangle_L+i|01\rangle_L+|10\rangle_L+i|01\rangle_L)/2$& & & & & & & & & 0.263& 0.230\\
 19&$(|00\rangle_L+|01\rangle_L+i|10\rangle_L+i|11\rangle_L)/2$& & & & & & & & & 0.287& 0.259 \\
 20&$(|00\rangle_L+i|01\rangle_L+i|10\rangle_L-|11\rangle_L)/2$& & & & & & & & & 0.275& 0.240\\
\hline
\hline
\end{tabular}
\caption{Distances of all experimental final states after nonadiabatic Holonomic quantum gates. Superscript $X1$ and $H1$ denote NOT and Hadamard gates in the single-loop way, superscript $X2$ and $H2$ denote NOT and Hadamard gates in the composite way, and superscript 2 denotes a two logical-qubit gate.  Subscript $F$ and $L$ denote the three-physical-qubit space and the logical-qubit subspace.}\label{SF1}
\end{table*}

\section{Quantum process tomography in the DFS}
In the maintext, we follow the standard  QPT~\cite{Chuang1997} method to experimentally reconstruct $\chi$ matrices in the logical qubit subspace for holonomic operations. The goal of QPT is to determine a fixed set of operation elements $\{\tilde E_i\}$ for a quantum channel $ \mathcal{E}$: $ \mathcal{E}(\rho) = \sum_{mn}  \chi_{mn} \tilde E_m \rho \tilde E_n$. Let $\rho_j$ $(1\le j \le d^2)$ be a fixed, linearly independent basis for the space of $ d \times d$ matrices.  Each $\mathcal{E}(\rho_j) $ may be expressed as a linear combination of the basis states $\mathcal{E}(\rho_j) = \sum_{k} \lambda_{j k} \rho_k $. Given that an input state $\rho_i$ and $\{\tilde E_i\}$ are known, one can determine the action of $\tilde E_m \rho_j \tilde E_n = \sum_k \beta_{jk}^{mn} \rho_k$, where $\beta_{jk}^{mn}$ are complex numbers which can be determined by standard algorithms. Thus $\sum_{k} \sum_{mn}  \chi_{mn} \beta_{jk}^{mn} \rho_k =  \sum_{k} \lambda_{j k} \rho_k $. From the linear independence of the $\rho_k$, it follows that $ \sum_{mn}  \beta_{jk}^{mn} \chi_{mn}  =   \lambda_{j k} $  for each $k$. Finally, one can determine $  \chi_{mn}$ given the known values for $\beta_{jk}^{mn} $ and $\lambda_{j k}$ using standard methods of linear algebra.

For single-logical-qubit gates, the fixed set of operation elements $\{\tilde E_i\}$ can be
\begin{eqnarray}
\tilde{E}_0=I_L, \tilde{E}_1=X_L, \tilde{E}_2=-iY_L, \tilde{E}_3=Z_L,
\end{eqnarray}
where $I_L =|0\rangle_L\langle 0| +|1\rangle_L\langle 1|  $, $X_L = |0\rangle_L\langle 1| + |1\rangle_L\langle 0|$, $Y_L = i |0\rangle_L\langle 1| - i |1\rangle_L\langle 0|$ and $Z_L = |0\rangle_L\langle 0| - |1\rangle_L\langle 1| $ in the DFS. There are 12 parameters, specified by $\chi_1$.
We prepare four input states as
 \begin{eqnarray}
\{|0\rangle_L,|1\rangle_L,(|0\rangle_L+|1\rangle_L)/\sqrt{2},(|0\rangle_L+i|1\rangle_L)/\sqrt{2}\}\nonumber,
\end{eqnarray}
and the final states through a quantum channel $\mathcal{E}$ are
\begin{eqnarray}
\rho'_1&=& \mathcal{E}(|0\rangle_L\langle 0|),\nonumber\\
\rho'_4&=& \mathcal{E}(|1\rangle_L\langle 1|),\nonumber\\
\rho'_2&=& \mathcal{E}(|+\rangle_L\langle +|)+i \mathcal{E}(|-\rangle_L\langle -|)-(1+i)(\rho'_1+\rho'_4)/2,\nonumber\\
\rho'_3&=& \mathcal{E}(|+\rangle_L\langle +|)-i \mathcal{E}(|-\rangle_L\langle -|)-(1-i)(\rho'_1+\rho'_4)/2,\nonumber
\end{eqnarray}
which can be reconstructed using quantum state tomography, i.e., experimental density matrices for three physical qubits. In order to illustrate the behaviors of quantum gates in the logical-qubit subspace, we only extract the matrix elements in the DFS $S_1$ to form $ \rho'^{L}_m = \sum_{i', j' \in \{100, 001 \}}  c_{i'j'} |i' \rangle \langle j' | $ from the three-qubit state $\rho'_m = \sum_{ i, j \in \{ 000 , 001, ....,  111\}} c_{ij} |i \rangle \langle j |$. The experimentally reconstructed results for the initial and final states in the three-physical-qubit space are respectively shown in Fig. \ref{FigureS3} and Fig. \ref{FigureS4}, where the elements in the logical-qubit subspace are marked as the dark bars. We calculate the corresponding distances of $ \rho'^{L}_m$ from the idea ones, listed in Table \ref{SF} and Table~\ref{SF1}. From the experimental $ \rho'^{L}_m$, we obtain the $\chi$ matrices for single logical-qubit gates in the DFS as
\begin{eqnarray}
\chi^L_1=\Lambda^L_1\left(
\begin{array}{cc}
\rho'^L_1 & \rho'^L_2\\
\rho'^L_3 & \rho'^L_4
\end{array}
\right)\Lambda^L_1,
\end{eqnarray}
with
\begin{eqnarray}
\Lambda_1&=&
\frac{1}{2}\left(
\begin{array}{cc}
 I_L & X_L  \\
 X_L & -I_L
\end{array}
\right).
\end{eqnarray}

For two-logical-qubit gates, we prepare  16 initial states $ | \psi_{nm} \rangle= |n \rangle \otimes |m \rangle$,
where $ |n \rangle, |m \rangle \in \{|0\rangle_{\rm{L}},|1\rangle_{\rm{L}},(|0\rangle_{\rm{L}}+|1\rangle_{\rm{L}})/\sqrt{2}, (|0\rangle_{\rm{L}}+i|1\rangle_{\rm{L}})/\sqrt{2}\}$, and measure the final states through the two-logical-qubit quantum channel: $\rho'_{mn} = \mathcal{E} (\rho_{mn} =| \psi_{nm} \rangle \langle \psi_{nm}| ) $. Like the case for single-logical-qubit gates, we reconstructed the physical-qubit state $\rho'_{mn}$ and then extract the elements in DFS $S_2$ to form $\rho'^L_{mn}$. From the experimental $\rho'^L_{mn}$, the $\chi$ matrices for two logical-qubit gates in DFS are achieved as
\begin{equation}
\chi_2=\Lambda^L_2\overline{\rho}'^L\Lambda^L_2,
\end{equation}
where $\Lambda^L_2=\Lambda^L_1\otimes\Lambda^L_1$, and
\begin{equation}
\overline{\rho}'^L=P^{T}_L
\left(
\begin{array}{ccccc}
 \rho'^L_{11}  & \rho'^L_{12}  & \rho'^L_{13}  & \rho'^L_{14} \\
 \rho'^L_{21}  & \rho'^L_{22}  & \rho'^L_{23}  & \rho'^L_{24} \\
 \rho'^L_{31}  & \rho'^L_{32}  & \rho'^L_{33}  & \rho'^L_{34} \\
 \rho'^L_{41} & \rho'^L_{42}  & \rho'^L_{43}  & \rho'^L_{44} \\
\end{array}
\right)P_L,
\end{equation}
where {  $P_L=I_L\otimes[(\rho^L_{11}+\rho^L_{23}+\rho^L_{32}+\rho^L_{44})\otimes I_L]$ and $P^{T}_L$ is the transposition of $P_L$.}

\section{Error analysis}

Table~\ref{SF} and Table~\ref{SF1} shows the distances of all initial states experimentally prepared in Fig. \ref{FigureS3} and Fig. \ref{FigureS4} from the idea ones. Inputting these experimental initial states to an idea quantum channel, the distance of the reconstructed $\chi$ matrices in the logical-qubit subspace by ideal QPT are around 0.148 and 0.167 for the single-logical qubit gates and two-logical-qubit gates, respectively. According to the experimental signal-to-noise ratio, we perform a numerical simulation by generating a white Gaussian noise on the measurements, which leads an error around 0.026. Consequently, the errors for the $\chi$-matrix QPT mainly come from the imperfection of the initial states, as well as that of quantum channel $\chi$ reconstructed, e.g., the nonadiabatic holonomic quantum gates.
We also find that the  gate infidelity for two-logical-qubit gates are larger than that of the single-logical-qubit cases. This is because that the two logical qubits have larger Hilbert subspace than that of the single-logical-qubit, and larger Hilbert space causes more errors involved in the matrix elements.

\end{document}